\documentclass[conference]{IEEEtran}
\IEEEoverridecommandlockouts
\usepackage{cite}
\usepackage{amsmath,amssymb,amsfonts}
\usepackage{algorithmic}
\usepackage{graphicx}
\usepackage{textcomp}
\usepackage{xcolor}
\usepackage{float}
\usepackage{enumitem}
\abovecaptionskip 
\belowcaptionskip 
\belowdisplayshortskip
\abovedisplayshortskip
\def\BibTeX{{\rm B\kern-.05em{\SemCom i\kern-.025em b}\kern-.08em
    T\kern-.1667em\lower.7ex\hbox{E}\kern-.125emX}}
\begin{document}

\title{Energy-Aware Service Offloading for Semantic Communications in Wireless Networks\\
}

\author{\IEEEauthorblockN{Hassan Saadat\textsuperscript{1}, Abdullatif Albaseer\textsuperscript{1}, Mohamed Abdallah\textsuperscript{1}, Amr Mohamed\textsuperscript{2}, Aiman Erbad\textsuperscript{1}\\
\textsuperscript{1} Division of Information and Computing Technology, College of Science and Engineering 
\\Hamad Bin Khalifa University, Doha, Qatar\\ 
\textsuperscript{2} Department of Computer Science and Engineering, Qatar University, Doha, Qatar
}
Email: \textsuperscript{1}\{hasa52143, aalbaseer, moabdallah, AErbad\}@hbku.edu.qa, \textsuperscript{2}amrm@qu.edu.qa}

\maketitle

\begin{abstract}
Today, wireless networks are becoming responsible for serving intelligent applications, such as extended reality and metaverse, holographic telepresence, autonomous transportation, and collaborative robots. Although current fifth-generation (5G) networks can provide high data rates in terms of Gigabytes/second, they cannot cope with the high demands of the aforementioned applications, especially in terms of the size of the high-quality live videos and images that need to be communicated in real-time. Therefore, with the help of artificial intelligence (AI)-based future sixth-generation (6G) networks, the semantic communication concept can provide the services demanded by these applications. Unlike Shannon's classical information theory, semantic communication urges the use of the semantics (meaningful contents) of the data in designing more efficient data communication schemes. Hence, in this paper, we model semantic communication as an energy minimization framework in heterogeneous wireless networks with respect to delay and quality-of-service constraints. Then, we propose a sub-optimal solution to the NP-hard combinatorial mixed-integer nonlinear programming problem (MINLP) by utilizing efficient techniques such as discrete optimization variables' relaxation. In addition, AI-based autoencoder and classifier are trained and deployed to perform semantic extraction, reconstruction, and classification services. Finally, we compare our proposed sub-optimal solution with different state-of-the-art methods, and the obtained results demonstrate its superiority.
\end{abstract}

\begin{IEEEkeywords}
semantic communication, 6G networks, classical information theory, energy minimization, resource allocation
\end{IEEEkeywords}

\section{Introduction}
In recent years, the purpose of data communication has transformed from being only concerned with transmitting text, images, videos, and live streaming into more complex and intelligent applications. For example: 1) virtual reality (VR), such as metaverse and VR games, 2) holograms, 3) intelligent transportation, such as self-driving cars, and 4) collaborative robots, such as in medical and industrial environments. These applications tend to communicate with very large sizes of live data and require very fast transmission in terms of milli or even microseconds. Although current fifth-generation (5G) networks can provide data rates of up to gigabytes/seconds, they still cannot cope with the requirements of the aforementioned highly-demanding intelligent applications. Therefore, a smarter and more efficient way of communication is urged to be deployed in future networks in order to serve modern applications. One of the promising solutions is to apply the concept of semantic communication (SemCom). 

The current traditional way of communication is based on Shannon's classical information theory (CIT)\cite{shannon}. In his CIT, Shannon believed that the semantics (meaningful contents) of the transmitted data should not affect the design of the communication framework. His belief aligns with the technical level of communication that Weaver defined in \cite{weaver}. SemCom, on the other side, tries to move beyond naive communication at the technical level and achieve higher levels of intelligent communication at the semantic and effectiveness levels. SemCom is a communication framework that opts to convert communicating nodes from being naive symbol-exchanging transmitters and receivers into becoming more intelligent agents that exchange the semantic parts of the data to help effectively and efficiently achieve the goal of communication \cite{less}. 

Deploying SemCom in heterogeneous wireless networks has many challenges that must be encountered. For example, although SemCom guarantees less transmission energy consumption compared to traditional communication -as less data is being transmitted-, the extra computational energy consumed during semantic extraction and reconstruction must be considered. Also, due to the heterogeneous nature of wireless networks, not all end devices can deploy the same semantic extraction techniques and respect the same delay constraints. In addition, SemCom introduces semantic extraction and reconstruction losses that have to be minimized. Furthermore, a semantic knowledge mismatch between the transmitter and the receiver could lead to errors that negatively affect the goal of communication.

To evaluate the capabilities of current networks and determine the required technologies for SemCom, Yang et al. \cite{semcom} present AI-enabled SemCom and its 6G key enablers, discuss its communication- and semantic-related techniques and challenges, and provide examples of applications where SemCom can be beneficial. In \cite{less}, Chaccour et al. provide a theoretical foundation for future SemCom. The authors view future SemCom as an intelligent reasoning-driven framework where communication sides build a common semantic language and knowledge base. DeepSC, a more practical work, is proposed by Xie et al. \cite{deepsc}. DeepSC sets the ground for numerous current deep learning (DL)-enabled SemCom implementations, where joint source-channel encoding is performed, a generalized transfer learning-based SemCom framework is proposed, and a new text semantic similarity metric is introduced.

The existing works in the literature have deployed SemCom in various applications. For example, Tong et al. \cite{audio} propose a federated learning-based audio transmission SemCom framework, and use convolutional neural network-based autoencoders for the semantic tasks. To apply the concept of disentangling the data sample into learnable and memorizable parts as discussed in their earlier work \cite{less}, Chaccour et al. \cite{disentangling} use a contrastive learning-based approach, which managed to reduce the size of transmitted data by 57\%, while maintaining the same semantic performance. In the work of Farshbafan et al., \cite{curriculum}, curriculum and reinforcement learning (RL) were used to build a common semantic language in a goal-oriented SemCom-based dynamic environment. In the paper of Wang et al. \cite{appo}, a RL-based solution is applied at the base station to optimize the resource allocation and semantic token selection ratio, such that maximal text reconstruction similarity at the users' sides is ensured with respect to delay constraints. 

Even though the aforementioned works tackle interesting SemCom-related problems, none of them takes into consideration the computation and transmission energy consumption of the system. Also, they either consider a system that contains one edge/base station and its connected users, or they assume distance-based user-edge association (i.e., user-edge association is not a decision variable in their optimization problem). In Yang et al.'s work \cite{rate}, a rate splitting multiple access (RSMA)-based \cite{RSMA} SemCom is deployed in downlink communication between a base station and its users. Here, the goal is to perform physical resource and semantic extraction ratio allocation to minimize the total energy consumption of the system with respect to a semantic accuracy threshold. Xiao et al. \cite{ifabric} apply DeepSC-based uplink SemCom in healthcare environments. In their system, medical users require intelligent health services, and the cloud performs user-edge association and bandwidth allocation, such that the total energy of the system is minimum and quality-of-service (QoS) is maximum. Noting that each edge, based on its available capabilities, can serve different services with different qualities.

As will be shown next, our work differs from the discussed works, such that we apply energy-aware image-based SemCom in wider networks and treat both the user-edge association and semantic extraction ratio allocation as part of our optimization decisions.

The objectives of this paper are highlighted as follows: 
\begin{enumerate}
    \item Model SemCom as an energy minimization framework in heterogeneous wireless networks, with respect to minimum QoS and maximum delay constraints. 
    \item Propose a sub-optimal solution to the NP-hard combinatorial mixed-integer nonlinear programming problem (MINLP) by utilizing efficient techniques such as discrete optimization variables' relaxation. 
    \item Train and deploy DL-based autoencoder and classifier to perform semantic extraction, reconstruction, and classification on the exchanged data. 
    \item Compare the proposed SemCom framework with state-of-the-art user-edge association methods, and the results show the superiority of our approach.
\end{enumerate}

\begin{figure*}[t]
\centering
\includegraphics[width=0.7\linewidth]{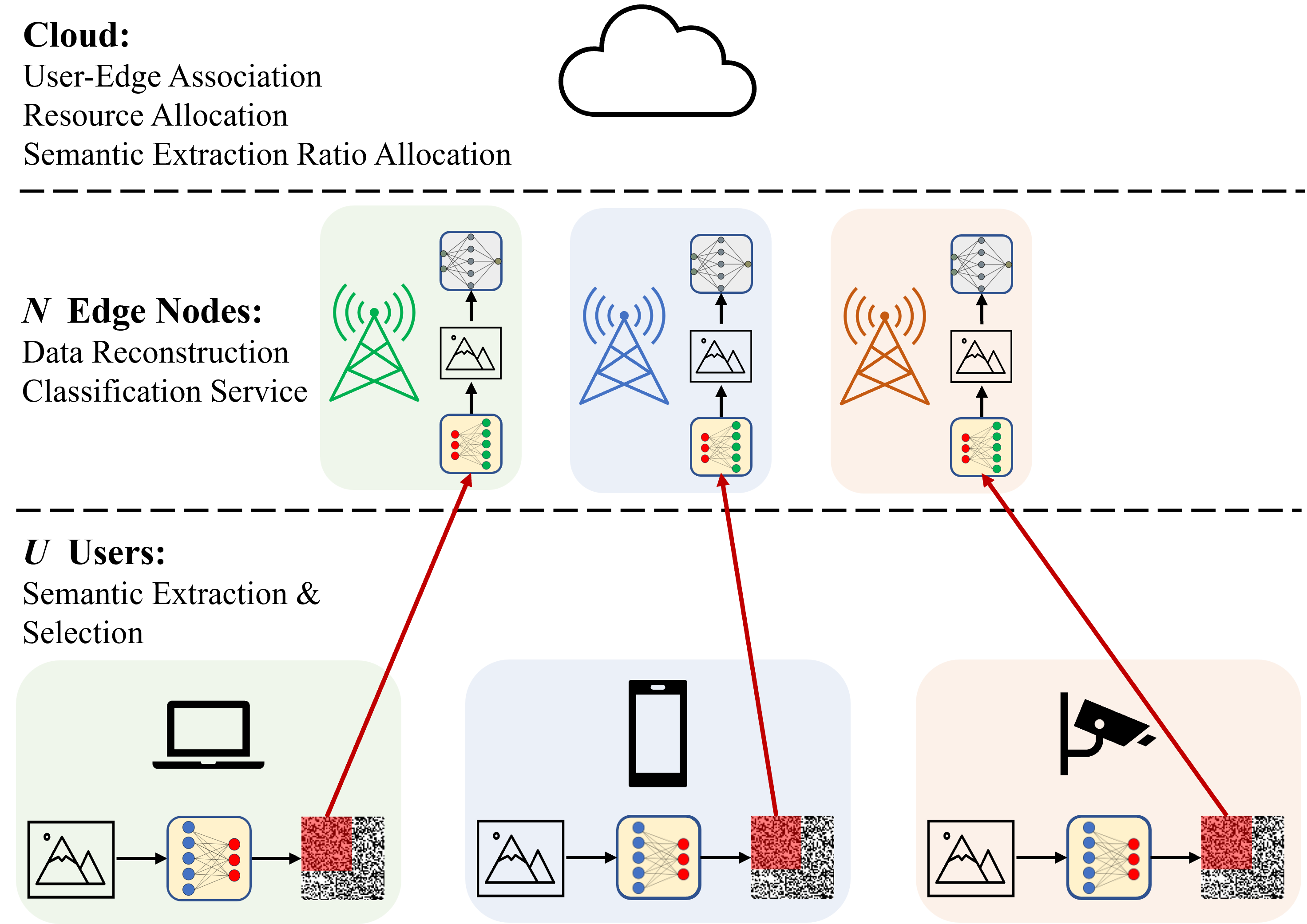}
\caption{System Model}
\label{sys_model}
\end{figure*}

\section{System Model \& Problem Formulation} 
In the system shown in Fig. \ref{sys_model}, we have a cloud server acting as a control unit, a set of edge nodes $N$ having a common semantic decoder with input size $\theta$ and providing a common service type (i.e., object classification), and a set of end users $U$ having a common semantic encoder with output size $\theta$ and requiring the service offered by the edge nodes. At each round, before the communication starts, the cloud performs user-edge association ($\mathbf{X}\in\{0,1\}$), semantic extraction ratio allocation ($\mathbf{\Delta}\in(0,1]$), user CPU frequency allocation ($\mathbf{F}\in\Re^+$), and edge CPU frequency allocation ($\mathbf{H}\in\Re^+$). The size of each decision matrix above is $|U|\times|N|$. 

Once the decision matrices are set, each user $u$ semantically encodes its image sample $D_u$ and randomly selects a fraction from the encoded image that is equal to the allocated $\delta_{un} \in (0,1]$, such that the total size of the data to be transmitted by each user $u$ becomes $\delta_{un} \times \theta$. Higher $\delta_{un}$ means that more semantic information is sent from user $u$ to edge $n$, which intuitively means edge $n$ can more accurately reconstruct user $u$'s image and perform the classification service. If we call the quality (classification accuracy) of service (QoS) provided by edge $n$ to user $u$ as $Q_{un}(\delta_{un})$, then the total QoS of the system can be calculated as:
\begin{equation}
    Q=\sum_{u=1}^{|U|}\sum_{n=1}^{|N|}x_{un}Q_{un}(\delta_{un}),
\label{Q}
\end{equation} where $x_{un}$ is the user-edge association variable, such that $x_{un}=1$ if user $u$ is associated with edge $n$, and $x_{un}=0$ otherwise.

The semantic extraction and selection operations by user $u$ take a computational time of:
\begin{equation}
    t_{un}^{comp_u}=\frac{y_{1un}(|D_u|,\theta,\delta_{un})}{f_{un}},
\label{tu}
\end{equation}and consume a computational energy calculated as:
\begin{equation}
    E_{un}^{comp_u}=\kappa y_{1un}(|D_u|,\theta,\delta_{un})f_{un}^2,
\label{Eu}
\end{equation} where $\kappa$ is the effective switched capacitance coefficient, $y_{1un}$ is the number of CPU cycles needed by user $u$ to encode its image and select $\delta_{un}$ part of the encoded image, and $f_{un}$ is the allocated local CPU frequency of user $u$ when associated with edge node $n$. 

Next, the transmission time needed to send the semantic information from user $u$ to edge $n$ is:
\begin{equation}
    t_{un}^{trans}=\frac{\delta_{un}\theta}{r_{un}},
\end{equation} 
where $r_{un}=b_{un}\log_2(1+\frac{p_{un}g_{un}}{b_{un}N_0})$ is the upload transmission rate, $b_{un}$ is the given bandwidth, $g_{un}$ is the channel gain between user $u$ and edge $n$, $N_0$ is the noise spectral density, and $p_{un}$ is the given transmission power for user $u$ to send its data to edge $n$. The transmission energy consumed can be calculated as: 
\begin{equation}
  E_{un}^{trans}=t_{un}^{trans}p_{un}=\frac{\delta_{un}\theta}{b_{un}\log_2(1+\frac{p_{un}g_{un}}{b_{un}N_0})}p_{un}.
\label{Et}
\end{equation}
After that, the computational time consumed by edge $n$ to reconstruct user $u$'s data and perform the classification service is calculated as:
\begin{equation}
t_{un}^{comp_e}=\frac{y_{2un}(|D_u|,\theta,\delta_{un})}{h_{un}}.
\label{te}
\end{equation}
and the computational energy is:
\begin{equation}
    E_{un}^{comp_e}=\kappa y_{2un}(|D_u|,\theta,\delta_{un})h_{un}^2,
\label{Ee}
\end{equation}where $y_{2un}$ is the number of CPU cycles required by edge $n$ to serve user $u$, and $h_{un}$ is the CPU frequency dedicated on edge $n$ to serve user $u$. 

The total energy consumed by the system can be calculated as:
\begin{equation}
    E=\sum_{u=1}^{|U|}\sum_{n=1}^{|N|}x_{un}(E_{un}^{comp_u}+E_{un}^{trans}+E_{un}^{comp_e}). 
\end{equation} 

To this end, we can define our goal as performing user-edge association, semantic extraction ratio allocation, user CPU frequency allocation, and edge CPU frequency allocation, such that the system's total energy is minimized, with respect to global minimum QoS and maximum delay thresholds (set by the cloud), as formulated in the following optimization problem:

\begin{equation}
\begin{aligned}
\min_{\mathbf{X,\Delta,F,H}} \quad E,
\end{aligned}
\label{p1}
\end{equation}
\textbf{s.t.}
\begin{equation}
x_{un}Q_{un}(\delta_{un})\geq Q^{min}, \quad \forall u\in{U}\: \& \:\forall n\in{N},
\label{qc} 
\end{equation}
\begin{equation}
\quad  x_{un}(t_{un}^{comp_u}+t_{un}^{trans}+t_{un}^{comp_e})\leq T^{max}, \quad  \forall u\in{U}\: \& \:\forall n\in{N},
\label{tc}
\end{equation}
\begin{equation}
x_{un}f_{un}\leq F_u^{max}, \quad  \forall u\in{U}\: \& \:\forall n\in{N},
\label{fc}
\end{equation}
\begin{equation}
\sum_{u=1}^{|U|}x_{un}h_{un}\leq H_n^{max}, \quad  \forall n\in{N},
\label{hc}
\end{equation}
\begin{equation}
    0<\delta_{un}\leq1, \quad  \forall u\in{U}\: \& \:\forall n\in{N},
\label{dc}
\end{equation}
\begin{equation}
0< f_{un},h_{un}, \quad  \forall u\in{U}\: \& \:\forall n\in{N},
\label{fhc}
\end{equation}
\begin{equation}
x_{un}\in\{0,1\}, \quad \forall u\in{U}\:\&\:\forall n\in{N},
\label{xc}
\end{equation}
\begin{equation}
\sum_{n=1}^{|N|}x_{un}=1, \quad  \forall u\in{U}.
\label{xcs}
\end{equation}
Constraint (\ref{qc}) ensures that each user gets served with a minimum QoS of $Q^{min}$.
Constraint (\ref{tc}) ensures that the maximum delay of the system is below $T^{max}$. Constraints (\ref{fc}) and (\ref{hc}) ensure that the allocated CPU frequencies of each user $u$ and edge node $n$ do not exceed their maximum capacity. Constraint (\ref{dc}) guarantees that the size of data to be sent by each user is within the range $(0\times\theta,1\times\theta]$. Constraint (\ref{fhc}) guarantees that each user gets served (gets allocated local and edge CPU frequencies). Finally, constraints (\ref{xc}) and (\ref{xcs}) allow each user $u$ to be associated with one edge node only.

\section{Proposed Solution}
To make our optimization problem in (\ref{p1}) convex and solvable, the following limitations must be encountered:
\begin{enumerate}[label=(\alph*)]
    \item The relationships between $Q_{un}$ and $\delta_{un}$ in (\ref{Q}), $y_{1un}$ and ($|D_u|,\theta,\delta_{un}$) in (\ref{Eu}), and $y_{2un}$ and ($|D_u|,\theta,\delta_{un}$) in (\ref{Ee}) need to be identified.
    \item Due to the binary user-edge association variable $x_{un}$, the optimization in (\ref{p1}) is a combinatorial mixed-integer nonlinear programming problem (MINLP) which is NP-hard \cite{Lee_Leyffer_2012}.
\end{enumerate} 

\begin{figure}[t]
\centering
\includegraphics[width=72mm]{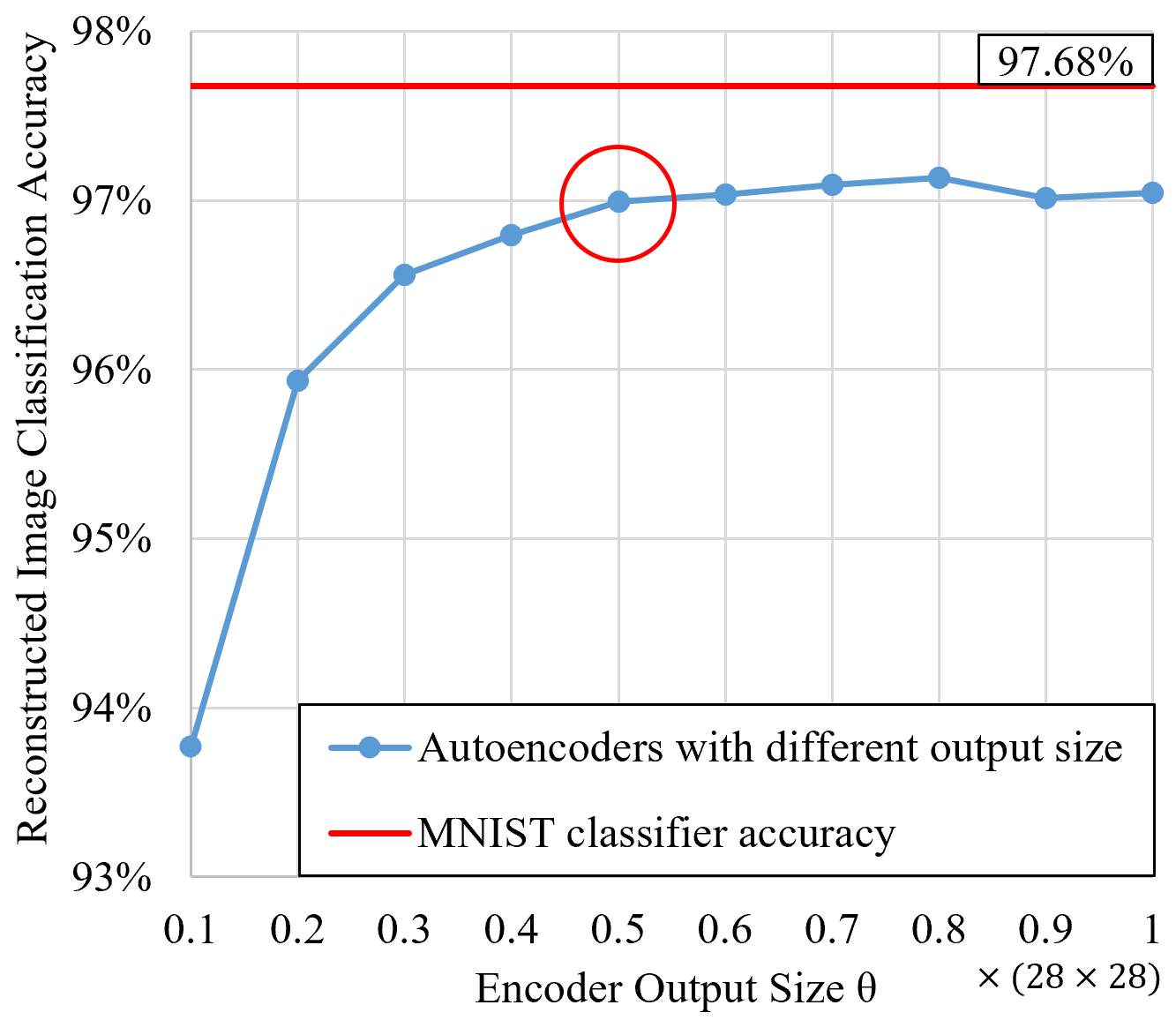}
\caption{Classification accuracy of reconstructed images by trained autoencoders with variable output size $\theta$.}
\label{Autoencoders_theta}
\end{figure}

\begin{figure}[H]
\centering
\includegraphics[width=72mm]{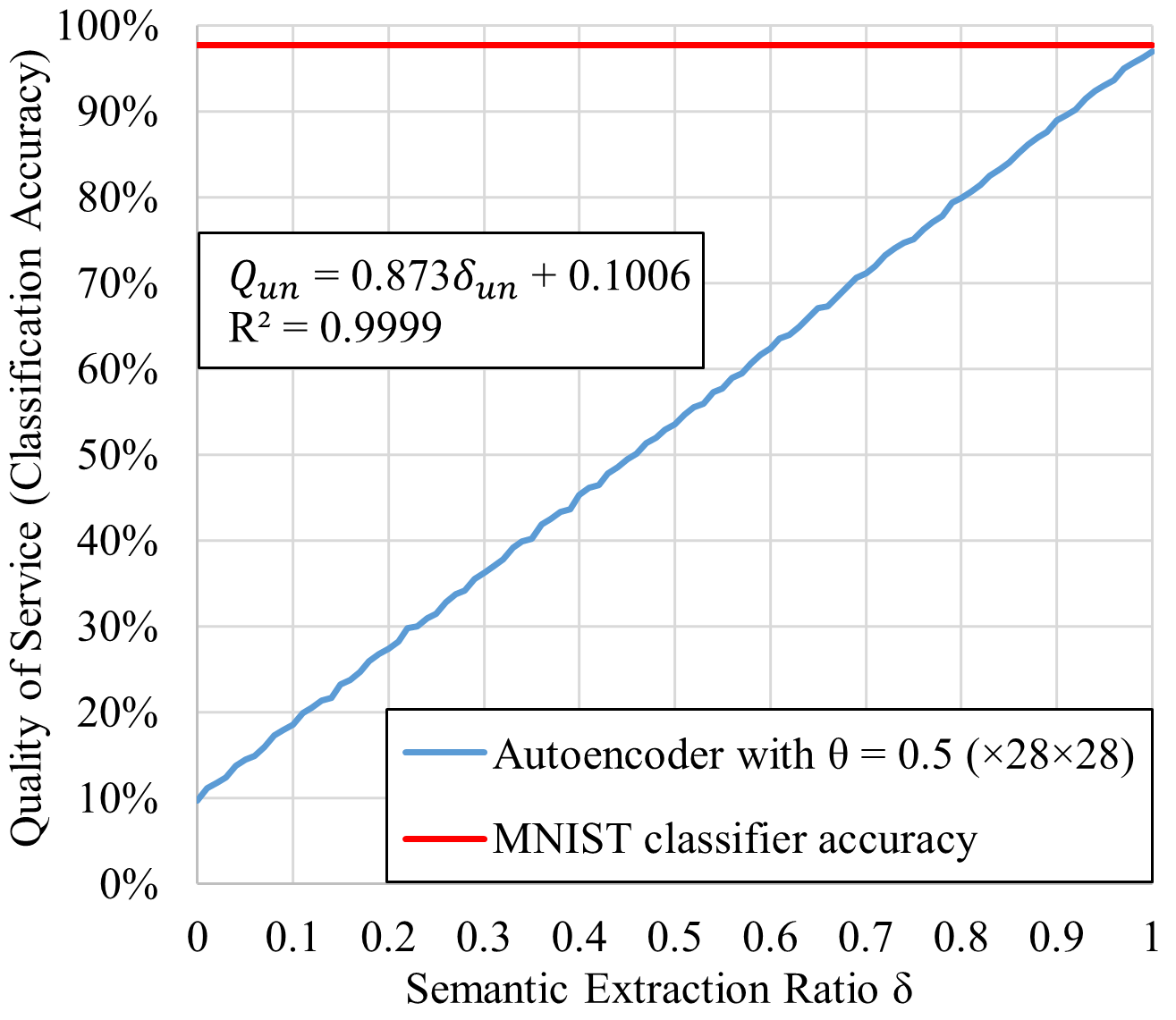}
\caption{Classification accuracy of reconstructed images by our autoencoder with $\theta = 0.5 (\times 28 \times 28)$ and variable semantic extraction ratio $\delta_{un}$.}
\label{Autoencoders_delta}
\end{figure}

To encounter the limitation mentioned in (a), first, we introduce the common autoencoder used in the system. On the users' side, we have an encoder that takes a $28\times 28$ image (MNIST image \cite{MNIST}) in the input layer and has a fully-connected output layer of size $\theta$. At each edge node's side, there exists a decoder that takes an encoded image of size $\theta$, and reconstructs it to its original size of $28\times 28$. Then, each edge node $n$ classifies the reconstructed image of its connected user(s) $u$ with a classification accuracy of $Q_{un}$ using a pre-trained MNIST classifier. The classifier has an input layer of $28\times28$ nodes, a hidden layer of 400 nodes, and an output layer of 10 nodes (labels). To select the encoder output size $\theta$, we train ten autoencoders for 5 epochs on the MNIST training dataset, each with a different value of $\theta$. From Fig. \ref{Autoencoders_theta}, we can see that the autoencoder's reconstruction classification accuracy starts converging to 97\% at $\theta=0.5(\times28\times28)$. Therefore, to reduce the size of the transmitted images on the network, it is safe to select and deploy the autoencoder with an output size of $\theta=0.5(\times28\times28)$ in our system. 

Then, to find the relationship between the QoS metric $Q_{un}$ and the semantic extraction ratio $\delta_{un}$, we encode the 10k-MNIST testing set using our selected encoder, then decode and classify the images based on a randomly selected $\delta_{un}$ portion of the encoded images. Note that regardless of the value of $\delta_{un}$, the decoder architecture is fixed and still expects an input size of $\theta$. Therefore, the receiver fills the missing $1-\delta_{un}$ portion with zero pixels. As can be seen in Fig. \ref{Autoencoders_delta}, a linearly perfect positive relationship between $Q_{un}$ and $\delta_{un}$ is observed, where $Q_{un}=0.873\delta_{un}+0.1006$. This relationship is because sending more semantic information to the receiver results in more accurate image reconstruction and classification. Based on this conclusion, (\ref{Q}) can be re-written as:
\begin{equation}
    Q=\sum_{u=1}^{|U|}\sum_{n=1}^{|N|}x_{un}(0.873\delta_{un}+0.1006)
\end{equation}

\begin{figure}[t]
\centering
\includegraphics[width=0.9\linewidth]{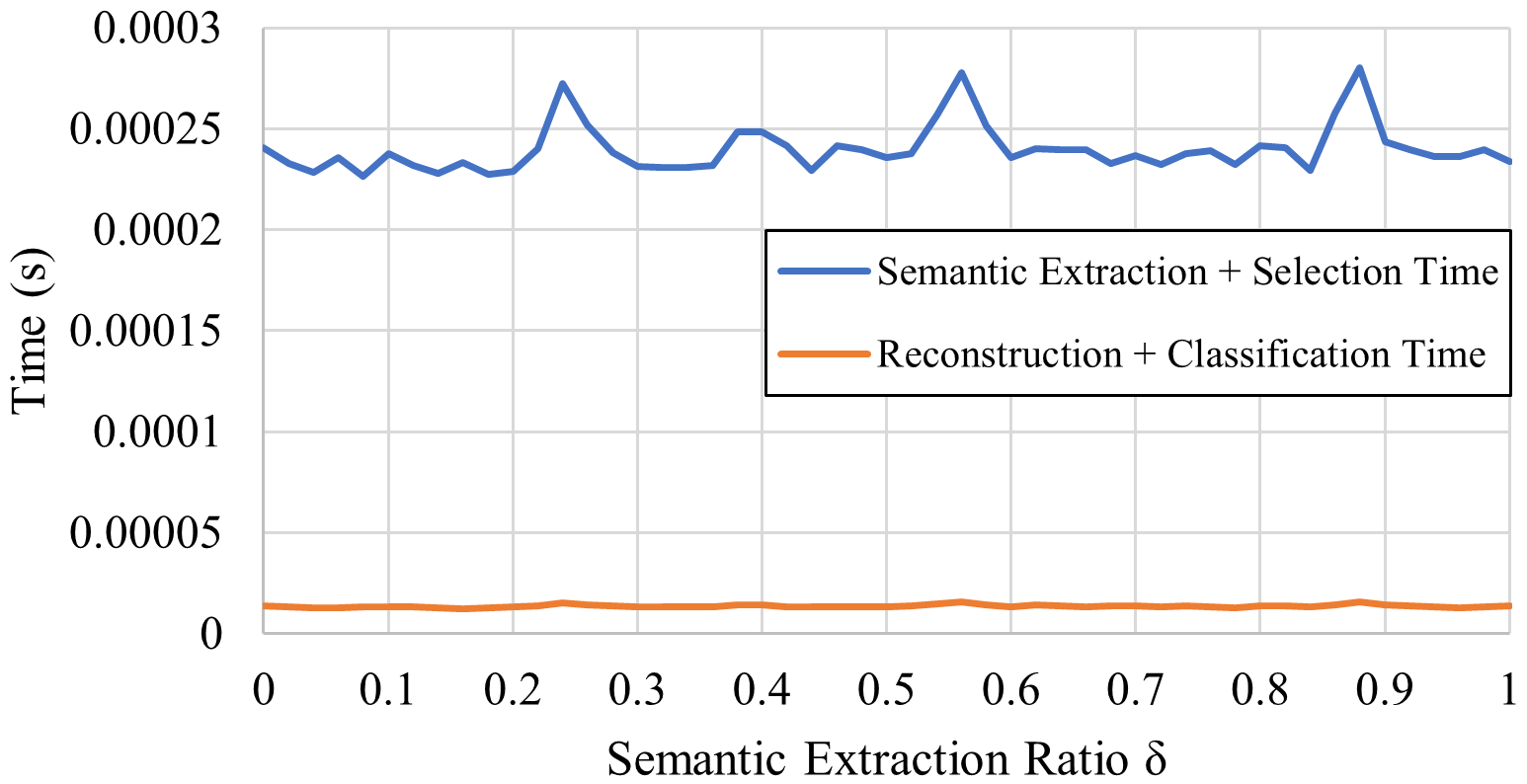}
\caption{Average computation time of semantic extraction and selection $t_{un}^{comp_u}$ at the user side, and reconstruction and classification $t_{un}^{comp_e}$ at the edge side.}
\label{Time}
\end{figure}

To identify the relationships $y_{1un}(|D_u|,\theta,\delta_{un})$ and $y_{2un}(|D_u|,\theta,\delta_{un})$, we need to bring our attention to the fact that $|D_u|$ in our system is fixed as $28\times28$ pixels (MNIST image), and $\theta$ is fixed as $0.5\times|D_u|$. Therefore, the number of CPU cycles $y_{1un}$ needed by user $u$ to perform semantic extraction and selection, and CPU cycles $y_{2un}$ needed by edge node $n$ to perform semantic reconstruction and classification, are only affected by the variable $\delta_{un}$. Now, to identify the relationships $y_{1un}(\delta_{un})$ and $y_{2un}(\delta_{un})$, we know from (\ref{tu}) and (\ref{te}) that for fixed CPU frequencies $f_{un}$ and $h_{un}$, the computation latencies $t_{un}^{comp_u}$ and $t_{un}^{comp_e}$ are directly proportional to $y_{1un}(\delta_{un})$ and $y_{2un}(\delta_{un})$, respectively. Therefore, we infer our autoencoder on the 10k-MNIST testing set, and measure $t_{un}^{comp_u}$ and $t_{un}^{comp_e}$ on different values of $\delta_{un}$. We can see in Fig. \ref{Time} that $t_{un}^{comp_u}$ and $t_{un}^{comp_e}$ are constants even with variable $\delta_{un}$. 
We can reason these findings to the fact that the $\delta_{un}$ portion selection happens after $D_{u}$ is encoded. Also, the edge nodes fill the missing $1-\delta_{un}$ portion with zero pixels and then perform reconstruction on $\theta$ pixels in all cases. 
Also, the $\delta_{un}$ portion is randomly selected with constant complexity. All in all, the same number of semantic computations are performed. Hence, $y_{1un}$ and $y_{2un}$ are constants and independent from $\delta_{un}$. 
The spikes in Fig. \ref{Time} represent negligible periodic operations on the experiment's host computer.

Finally, to sub-optimally solve the combinatorial MINLP introduced in (b), we relax the user-edge association matrix $\textbf{X}$ to be any real number $\in [0,1]$. Then, after the sub-optimal problem is solved, each user $u$ is associated with edge node $n$ that corresponds to the highest $x_{un}$. 

Now that we have countered the limitations defined in (a) and (b), we can reformulate our original optimization problem in (\ref{p1}) into a sub-optimal problem as follows:
\begin{equation}
\begin{aligned}
\min_{\mathbf{X,\Delta,F,H}} \quad \sum_{u=1}^{|U|}\sum_{n=1}^{|N|}x_{un}( \kappa y_{1un}f_{un}^2+\\ \frac{\delta_{un}\theta}{b_{un}\log_2(1+\frac{p_{un}g_{un}}{b_{un}N_0})}p_{un}+\kappa y_{2un}h_{un}^2),
\end{aligned}
\label{p2}
\end{equation}
\textbf{s.t.}
\begin{equation}
\frac{Q^{min}-0.1006}{0.873}\leq \delta_{un} \leq 1, \quad  \forall u\in{U}\: \& \:\forall n\in{N},
\label{qc2} 
\end{equation}
\begin{equation}
\quad  t_{un}^{comp_u}+t_{un}^{trans}+t_{un}^{comp_e}\leq T^{max}, \quad  \forall u\in{U}\: \& \:\forall n\in{N},
\label{tc2}
\end{equation}
\begin{equation}
0\leq f_{un}\leq F_u^{max}, \quad  \forall u\in{U}\: \& \:\forall n\in{N},
\label{fc2}
\end{equation}
\begin{equation}
\sum_{u=1}^{|U|}h_{un}\leq H_n^{max}, \quad  \forall n\in{N},
\label{hc2}
\end{equation}
\begin{equation}
0\leq h_{un}, \quad  \forall u\in{U}\: \& \:\forall n\in{N},
\label{fhc2}
\end{equation}
\begin{equation}
x_{un}\in[0,1], \quad  \forall u\in{U}\:\&\:\forall n\in{N},
\label{xc2}
\end{equation}
\begin{equation}
\sum_{n=1}^{|N|}x_{un}=1, \quad  \forall u\in{U}.
\label{xcs2}
\end{equation}

The sub-optimal problem in (\ref{p2}) is convex and is an instance of geometric programming optimization problems, and can be solved using the GPkit library on python \cite{gpkit}. As seen in constraint (\ref{qc2}), $\delta_{un}$ boundaries are redefined by utilizing the relationship found in Fig. \ref{Autoencoders_delta}. Also, $x_{un}$ is taken out from constraints (\ref{tc2}), (\ref{fc2}), and (\ref{hc2}), in order to guarantee that these constraints are not violated after performing de-relaxation on $x_{un}$. Furthermore, we can see the relaxation of the user-edge association variable $x_{un}$ in (\ref{xc2}).

\section{Performance Evaluation}
In this section, we compare the performances of three different user-edge association methods: 1) our proposed sub-optimal solution as in (\ref{p2}), 2) minimum distance-based user-edge association, and 3) random user-edge association. In the minimum distance and random associations, problem (\ref{p2}) is solved to determine $\textbf{F}$, $\textbf{H}$, and $\mathbf{\Delta}$, while $\textbf{X}$ is treated as a given parameter and is no longer an optimization variable. For the sake of comparison, we run two experiments: 1) measuring the total energy consumption with variable delay threshold $T_{max}$, and 2) measuring the total energy consumption with variable QoS threshold $Q_{min}$.

\subsection{Simulation settings}
In our experiments, we consider a system with $|U|=20$ users and $|N|=4$ edge nodes. The users are randomly distributed in a circle of 500 meter radius, and the edge nodes are randomly located in a ring with an inner radius of 500 meters and an outer radius of 1000 meters of the same center \cite{ifabric}. Each user $u$ has a maximum local CPU frequency capacity of $F_u^{max}\in \{0.5, 1.0, 1.5, 2.5\}$ GHz. All edge nodes have the same maximum CPU frequency capacity, that is, $H_n^{max}= 4$ GHz, $\forall n\in N$. The effective switched capacitance coefficient is set as $\kappa=10^{-28}$ \cite{rate}. As for the network resources, each edge node can allocate a total transmission power of $P_n^{max}=30$ dBm to the users and a total bandwidth of $B_n^{max}=20$ MHz. The transmission power and bandwidth of each edge node $n$ are divided among all users in fixed values proportional to their distance $d_{un}$.

\subsection{Energy Consumption with Variable $T^{max}$}
\begin{figure}[t]
\centering
\includegraphics[width=62mm]{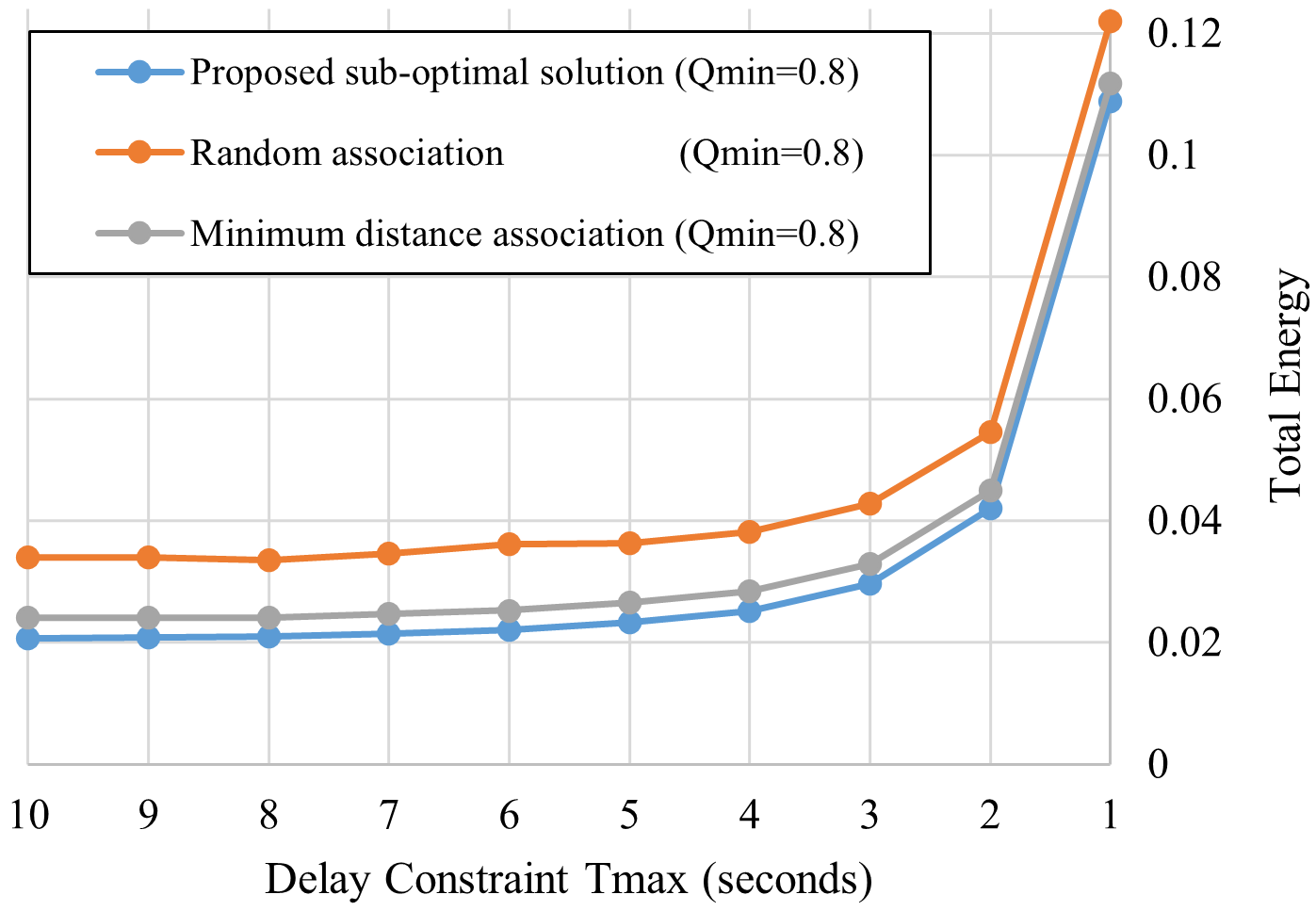}
\caption{Total system energy consumption with variable threshold $T^{max}$}
\label{Energy_T}
\end{figure}

In the experiment shown in Fig. \ref{Energy_T}, we measure the system's total energy consumption at a fixed minimum QoS threshold $Q^{min}=0.8$ while varying the maximum delay threshold $T^{max}=1,2,\dots,10$ seconds. Since $Q^{min}$ is fixed at 0.8, we know from the relationship found in Fig. \ref{Autoencoders_delta} that the lower bound of $\delta_{un}$ should be $\frac{0.8-0.1006}{0.873}=0.8011\leq \delta_{un}$. 
Also, from (\ref{p2}), in order to minimize the transmission energy, the optimization decision will allocate the minimum allowed $\delta_{un}$ to all users that guarantee to achieve constraint (\ref{qc2}), which is 0.8011. 
In other words, for a fixed $Q^{min}=0.8$ and variable $T^{max}$, the transmission energy $E^{trans}_{un}$ and transmission time $t^{trans}_{un}$ stay constant as the amount of data to be transmitted is fixed as $0.8011\times\theta$. Therefore, as shown in Fig. \ref{Energy_T}, as the time constraint gets stricter (as $T^{max}$ decreases), the optimization decision needs to allocate more CPU frequencies (higher $f_{un}$ and $h_{un}$) in order to reduce $t^{comp_u}_{un}$ and $t^{comp_e}_{un}$, which results in increasing the user and edge computational energy consumption by a factor of $f_{un}^2$ and $h_{un}^2$, respectively.

\subsection{Energy Consumption with Variable $Q^{min}$}
\begin{figure}[t]
\centering
\includegraphics[width=62mm]{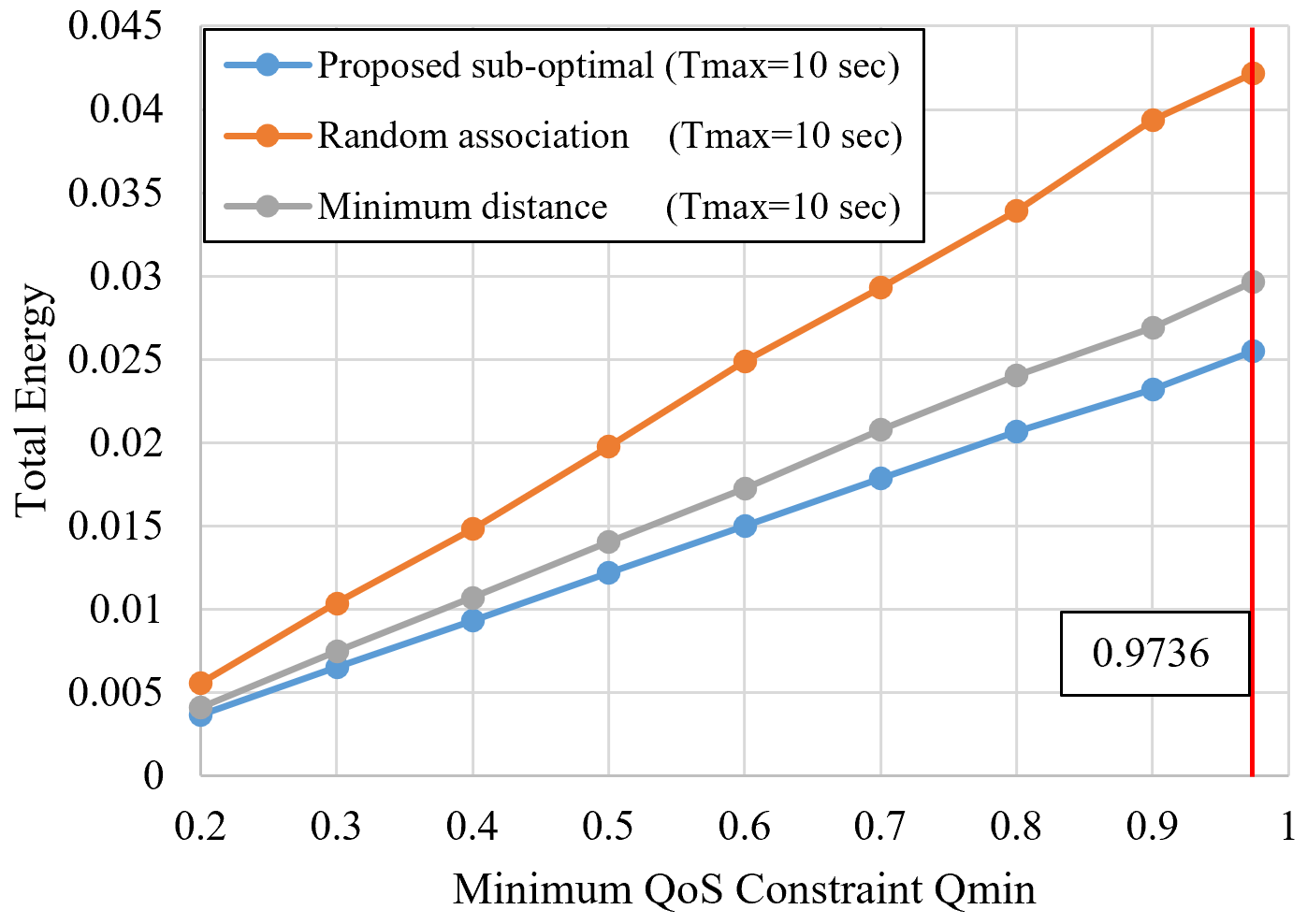}
\caption{Total system energy consumption with variable threshold $Q^{min}$}
\label{Energy_Q}
\end{figure}

In the experiment shown in Fig. \ref{Energy_Q}, we measure the system's total energy consumption at a fixed maximum delay threshold of $T^{max}=10$ seconds while varying the minimum QoS threshold $Q^{min}=0.2,0.3,\dots,0.9,0.9736$ (maximum possible $Q_{un}$ when $\delta_{un}=1$). Since $Q^{min}$ increases, the lower bound of $\delta_{un}$ also increases. And we know from Fig. \ref{Time} that the user and edge computational latencies ($t^{comp_u}_{un}$ and $t^{comp_e}_{un}$) are independent of the value of $\delta_{un}$ and $Q^{min}$, which also means that the user and edge computational energy consumption ($E^{comp_u}_{un}$ and $E^{comp_e}_{un}$) are independent of the value of $Q^{min}$. Therefore, we conclude that the increase in the total energy as a result of increasing $Q^{min}$ (Fig. \ref{Energy_Q}) is solely reasoned to an increase in the transmission energy $E^{trans}_{un}$. Also, the linearity can be reasoned to the fact that the transmission energy is linearly and directly proportional to the size of data to be transmitted ($\delta_{un}\theta $), i.e., $E^{trans}_{un}\propto\delta_{un}$, which is clearly shown in equation (\ref{Et}).

In both experiments, we can see that our sub-optimal solution outperforms the other baselines as it consumes less energy. The distance-based association provides a competitive performance. The reason behind that is, as stated earlier, the only decision that differentiates between the performances of the user-edge association methods is $\textbf{X}$, while all the other decision matrices $\textbf{F}$, $\textbf{H}$, and $\mathbf{\Delta}$, are determined by solving the sub-optimal problem (\ref{p2}). In other words, the differences in the total energy consumption is much more significantly affected by the transmission energy than the computational energy. Therefore, the minimum distance association method performance is close to the sub-optimal solution because a shorter distance between communication sides generally consumes less transmission energy and, hence, less total energy consumption. Finally, the random association consumes the highest total energy because it does not consider the distances and channel states between the users and edge nodes.

\section{Conclusion}
In this paper, the SemCom concept was briefly introduced, its deployment challenges were concisely discussed, and several SemCom-related literature works were reviewed. Then, an energy minimization SemCom framework that considers delay and QoS constraints was modeled. In addition, a DL-based autoencoder and classifier trained on the MNIST dataset were deployed for the semantic task. After that, regression and relaxation techniques were utilized to propose a sub-optimal solution to the formulated combinatorial MINLP optimization problem. Next, comparison experiments between the proposed sub-optimal solution and baseline user-edge association methods at different delay and QoS constraints were conducted. The results show that the proposed sub-optimal solution efficiently managed to reduce the energy consumption of the system while maintaining the defined constraints, compared to the other baseline methods. Finally, further enhancements can be done in future work, such as 1) making the solution more abstract and generalizable rather than being specific to our defined application, and 2) adding the transmission power and bandwidth to the optimization problem as decision variables.

\section{Acknowledgment}
This work was made possible by GSRA grant \# GSRA9-L-1-0519-22025 from the Qatar National Research Fund (a member of Qatar Foundation). The findings achieved herein are solely the responsibility of the authors.

\bibliography{refs}

\begin{thebibliography}{10}
\providecommand{\url}[1]{#1}
\csname url@samestyle\endcsname
\providecommand{\newblock}{\relax}
\providecommand{\bibinfo}[2]{#2}
\providecommand{\BIBentrySTDinterwordspacing}{\spaceskip=0pt\relax}
\providecommand{\BIBentryALTinterwordstretchfactor}{4}
\providecommand{\BIBentryALTinterwordspacing}{\spaceskip=\fontdimen2\font plus
\BIBentryALTinterwordstretchfactor\fontdimen3\font minus \fontdimen4\font\relax}
\providecommand{\BIBforeignlanguage}[2]{{%
\expandafter\ifx\csname l@#1\endcsname\relax
\typeout{** WARNING: IEEEtran.bst: No hyphenation pattern has been}%
\typeout{** loaded for the language `#1'. Using the pattern for}%
\typeout{** the default language instead.}%
\else
\language=\csname l@#1\endcsname
\fi
#2}}
\providecommand{\BIBdecl}{\relax}
\BIBdecl

\bibitem{shannon}
C.~E. Shannon, ``A mathematical theory of communication,'' \emph{The Bell System Technical Journal}, vol.~27, no.~3, pp. 379--423, 1948.

\bibitem{weaver}
\BIBentryALTinterwordspacing
W.~WEAVER, ``Recent contributions to the mathematical theory of communication,'' \emph{ETC: A Review of General Semantics}, vol.~10, no.~4, pp. 261--281, 1953. [Online]. Available: \url{http://www.jstor.org/stable/42581364}
\BIBentrySTDinterwordspacing

\bibitem{less}
C.~Chaccour, W.~Saad, M.~Debbah, Z.~Han, and H.~V. Poor, ``Less data, more knowledge: Building next generation semantic communication networks,'' 2022.

\bibitem{semcom}
W.~Yang, H.~Du, Z.~Q. Liew, W.~Y.~B. Lim, Z.~Xiong, D.~Niyato, X.~Chi, X.~Shen, and C.~Miao, ``Semantic communications for future internet: Fundamentals, applications, and challenges,'' \emph{IEEE Communications Surveys \& Tutorials}, vol.~25, no.~1, pp. 213--250, 2023.

\bibitem{deepsc}
\BIBentryALTinterwordspacing
H.~Xie, Z.~Qin, G.~Y. Li, and B.-H. Juang, ``Deep learning enabled semantic communication systems,'' \emph{{IEEE} Transactions on Signal Processing}, vol.~69, pp. 2663--2675, 2021. [Online]. Available: \url{https://doi.org/10.1109/tsp.2021.3071210}
\BIBentrySTDinterwordspacing

\bibitem{audio}
H.~Tong, Z.~Yang, S.~Wang, Y.~Hu, W.~Saad, and C.~Yin, ``Federated learning based audio semantic communication over wireless networks,'' in \emph{2021 IEEE Global Communications Conference (GLOBECOM)}, 2021, pp. 1--6.

\bibitem{disentangling}
C.~Chaccour and W.~Saad, ``Disentangling learnable and memorizable data via contrastive learning for semantic communications,'' 2022.

\bibitem{curriculum}
M.~K. Farshbafan, W.~Saad, and M.~Debbah, ``Curriculum learning for goal-oriented semantic communications with a common language,'' \emph{IEEE Transactions on Communications}, vol.~71, no.~3, pp. 1430--1446, 2023.

\bibitem{appo}
Y.~Wang, M.~Chen, T.~Luo, W.~Saad, D.~Niyato, H.~V. Poor, and S.~Cui, ``Performance optimization for semantic communications: An attention-based reinforcement learning approach,'' \emph{IEEE Journal on Selected Areas in Communications}, vol.~40, no.~9, pp. 2598--2613, 2022.

\bibitem{rate}
Z.~Yang, M.~Chen, Z.~Zhang, and C.~Huang, ``Energy efficient semantic communication over wireless networks with rate splitting,'' \emph{IEEE Journal on Selected Areas in Communications}, vol.~41, no.~5, pp. 1484--1495, 2023.

\bibitem{RSMA}
Y.~Mao, O.~Dizdar, B.~Clerckx, R.~Schober, P.~Popovski, and H.~V. Poor, ``Rate-splitting multiple access: Fundamentals, survey, and future research trends,'' \emph{IEEE Communications Surveys \& Tutorials}, vol.~24, no.~4, pp. 2073--2126, 2022.

\bibitem{ifabric}
W.~Xiao, Y.~Tang, J.~Liu, D.~Wu, B.~Alzahrani, Y.~Hao, and N.~Zhou, ``Semantic-driven efficient service network towards smart healthcare system in intelligent fabric,'' \emph{IEEE Transactions on Network Science and Engineering}, pp. 1--10, 2022.

\bibitem{Lee_Leyffer_2012}
J.~Lee and S.~Leyffer, \emph{Mixed integer nonlinear programming}.\hskip 1em plus 0.5em minus 0.4em\relax Springer, 2012.

\bibitem{MNIST}
L.~Deng, ``The mnist database of handwritten digit images for machine learning research,'' \emph{IEEE Signal Processing Magazine}, vol.~29, no.~6, pp. 141--142, 2012.

\bibitem{gpkit}
E.~Burnell, N.~B. Damen, and W.~Hoburg, ``\hbox{GPkit}: A human-centered approach to convex optimization in engineering design,'' in \emph{Proceedings of the 2020 {CHI} Conference on Human Factors in Computing Systems}, 2020.

\end{thebibliography}
\bibliographystyle{IEEEtran} 
\end{document}